\documentclass[useAMS,usegraphicx]{mn2e}
\usepackage{url}
\newcommand{\Msun}{\mbox{$\rm M_{\odot}$}}

\title[Carbon deficiencies in some classical Algols ]
      {Carbon deficiencies in the primaries of some classical Algols\thanks{Based on spectroscopic observations collected at Asiago (Italy) and T\"{U}B\.{I}TAK National Observatory (Turkey).}}

\author[\.{I}bano\v{g}lu et al.]
{ C.~\.{I}bano\v{g}lu$^1$, A.~Dervi{\c s}o{\u g}lu$^1$\thanks{Coressponding author: ahmetdervisoglu@mail.ege.edu.tr}, \"{O}. \c{C}ak{\i}rl{\i}$^1$,  E.~Sipahi$^1$ and K. Y\"{u}ce$^2$  \\
$^1$Ege University, Science Faculty, Department of Astronomy and Space Sciences, 35100 Bornova, \.{I}zmir, Turkey\\
$^2$Ankara University, Science Faculty, Department of Astronomy and Space Sciences, 06100 Tando\v{g}an, Ankara, Turkey\\
}

\begin{document}

\date{Accepted 2011 Month Day. Received 2011 Month Day; in original form 2011 ??? ?}
\pagerange{\pageref{firstpage}--\pageref{lastpage}} \pubyear{2011}
\maketitle
\label{firstpage}

\begin{abstract}

   The equivalent widths of C II $\lambda$ 4267 \AA~ line were measured for the mass-gaining primary stars of the 18 Algol-type binary systems. The comparison of the EWs of the gainers with those of the single standard stars having the same effective temperature and luminosity class clearly indicates that they are systematically smaller than those of the standard stars. The primary components of the classical Algols, located in the main-sequence band of the HR diagram, appear to be C poor stars. We estimate $ [N_{C} /N_{tot}] $ relative to the Sun as -1.91 for GT Cep, -1.88 for AU Mon and -1.41 for TU Mon, indicating poorer C abundance.  An average differential  carbon abundance has been estimated to be -0.82 dex relative to the Sun and -0.54 dex relative to the main-sequence standard stars. This result is taken to be an indication of the transferring material from the evolved less-massive secondary components to the gainers such that the CNO cycle processed material changed the original abundance of the gainers. There appear to be relationships between the EWs of C II $\lambda$ 4267 \AA~line and the rates orbital period increase and mass transfer in some Algols. As the mass transfer rate increases the EW of the C II line decreases, which indicates that accreted material has not been completely mixed yet in the surface layers of the gainers.  This result supports the idea of mixing as an efficient process to remove the abundance anomaly built up by accretion. Chemical evolution of the classical Algol-type systems may lead to constrains on the initial masses of the less massive, evolved, mass-losing stars.      
\end{abstract}

\begin{keywords}
stars:abundances-stars:atmospheres-stars:binaries:close\\-stars:binaries:eclipsing
\end{keywords}

\section{introduction}
The classical Algol-type binaries are composed of a B or A spectral-type main-sequence hotter primary star and an F or later-G giant or subgiant cooler secondary star. They are semi-detached interacting binary systems in which the evolved less massive secondary components have filled their corresponding Roche lobes. Therefore, the material on the less massive secondary star is transferring onto the hot main-sequence primaries. The classical Algols are, in general, in a slow stage of mass transfer. The evolution of the semi-detached binaries is mostly interpreted in the framework of the Roche model. A number of implicit assumptions are made for its application. These assumptions are ordered as follows: 1) the components are point masses, they are co-rotating with the orbital motion and 2) the most importantly the total mass and angular momentum are conserved.   However, these restrictions are not so valid for the evolved binary systems. Glazunova et al. (2008) measured rotational velocities of the primary stars in the classical Algols. At least seven systems out of 23 binaries rotate faster which differ from synchronize rotation by more than a factor of two. Eggleton (2000) discussed evolution of Algol type systems and concluded that they could not be evolved to their present status without having lost substantial mass and angular momentum.The fundamental parameters of well-observed detached and semi-detached Algols are compiled and analysed by Ibano{\u g}lu et al. (2006) to reveal some possible implications for their nuclear and angular momentum evolution. They arrived at a  result that the mass-ratio of detached Algols is larger than unity. As the system evolved-off from the main-sequence mass transfer begins and the evolution will proceed towards lower mass ratios without considerable angular momentum loss (AML). When the mass-ratio is reversed and became smaller than about 0.4 the orbital AML rate increases. It is also indicated that the classical Algols are separated into two subclasses with respect to their orbital periods, i.e. P$>$5 and P$<$5 days.  
  
Recently,  Dervi{\c s}o{\u g}lu et al. (2010) re-discussed spin angular momentum evolution of the long period Algols. They have demonstrated that even a small amount of mass transfer, gainer immediately spin up to the critical rotational velocity. However, the observed rotational velocities of gainers are smaller than 40 per cent of the critical rate. They considered generation of magnetic fields in the radiative atmospheres in a differentially rotating star and proposed the possibility of mass and angular momentum loss driven by strong stellar winds in the intermediate-mass stars, similar to the primaries of the Algols. The slow rotation of the primaries in the Algol systems is explained by a balance between the spin-up by mass accretion and spin-down by stellar wind linked to a magnetic field. Moreover, it is indicated that larger mass loss from the system is produced in the smaller magnetic fields.

For the first time, Parthasarathy et al. (1983) called attention to the carbon deficiencies and nitrogen over-abundances in the atmospheres of secondary components of U Cep and U Sge. Later on,  Cugier and Hardorp (1988) indicated the carbon deficiencies in the IUE-spectra of the gainers in $\beta$ Per (Algol) and $\lambda$ Tau. Cugier (1989) expanded his study on the Algols using the IUE archival data and found similar results for six stars as in the case of Algol and $\lambda$ Tau. Tomkin and his collaborators (Tomkin, Lambert and Lemke 1993) observed eight Algols in the optical wavelengths and compared the C abundances in the primaries with those at the single standard stars having nearly the same effective temperatures and luminosities. They arrived at a result that the mass-gaining primaries of the semi-detached binaries have smaller C abundance with respect to the standard stars. On the other hand Yoon and Honeycutt (1992) measured C abundance for 12 Algol secondaries using the strength of $g$-band of the CH molecule. The values of $\log\,\varepsilon(C)$ for the sample are smaller about 0.25-0.75 dex  than those field G and K giants.    

The distribution of C, N, O elements in the hydrogen-burning core of initially more massive component of an Algol type system  has been changed during the main-sequence evolution. In the Case B evolution of the close binary systems the more massive star expands and fills its Roche lobe as well as develops convection. Convective mixing in the atmosphere may change the distribution of the C and N.  However, carbon determinations for the primary and also for the secondary stars of Algol-type binaries appear to be insufficient to arrive at a relevant quantitative analysis and tests for accretion and mixing. Determinations of the carbon abundance for a large sample of Algols may act as major constraints on the evolution models for these systems (Sarna and de Greve 1996, 1997).

In this study, the results of spectroscopic observations of some Algols are presented. The equivalent width (EW) of C II $\lambda$ 4267 \AA~line was measured for the 18 systems. The differences of EWs between the Algol primaries and the standard stars having similar effective temperatures were determined and compared with the orbital period increase and mass-transfer rates. We present the carbon deficiencies for the largest sample of Algol-type mass-transferring systems and find, for the first time, that an evidence of a possible relationship between the carbon deficiency and mass transfer rate, at least for some systems which show orbital period increase.

\section{Observations}
The chemical abundance determinations from equivalent width analysis for the mass-losing secondary stars in the semi-detached Algol-type binaries could only be made  during the totality, when the more massive primary star is completely   eclipsed. Out of the primary eclipse, the light contribution of the donors does not exceed a few per cent. During the primary eclipse the brightness of these systems are too low to be taken a spectrum in the totality, requiring a large telescope and an appropriate  spectrograph. Therefore we prefer to take spectrum of the gainers which are dominate in the spectra and have enough effective temperatures that the lines of ionized carbon and nitrogen can be formed. However, the primaries of the Algols rotate fast enough that the blending affects the spectral lines. The gainers in the classical Algols rotate at least five or more than that synchronous rotation. Therefore a few lines of the carbon species can be measured. 

Spectroscopic observations were carried out at two sites, namely, Asiago and Turkish National observatories. The targets were selected to the capability of the instruments. At the Asiago Observatory (ASI) the selected systems were observed with the REOSC Echelle spectrograph and CCD mounted at Cassegrain focus of the 182 cm telescope. The spectra cover the wavelength interval between 3900 and 7300 \AA, divided into 27 orders. The average signal-to-noise ratio (S/N) and resolving power $\lambda$/$\Delta \lambda$ were about $\sim$ 150 and 50\,000, respectively. The observations were made between 10 and 20 March 2009 on successive nine nights. During this time interval 45 spectra of 14 Algols and 3 spectra of the three standard stars were obtained.

 In the spectroscopic observations at the Turkish National Observatory (TUG) the Coude Echelle Spectrometer  (CES) attached to the 150 cm telescope was used\footnote{Further details on the 
telescope and the spectrograph can be found at \url{http://www.tug.tubitak.gov.tr}.}. The wavelength coverage of each 
spectrum was 3700-10000 \AA~ in 85 orders, with a resolving power of $\lambda$/$\Delta \lambda \sim$ 125\,000 at 
4267 \AA~ and an average signal-to-noise ratio was $\sim$150. The observations were obtained on 28 and 29 May, 2010. During two nights observations 7 spectra of five Algols and 4 spectra of the three standard stars were obtained.
     
The position of the grating was chosen so that the C II $\lambda$ 4267 \AA~line was recorded simultaneously in the 9th and 10th orders with the H$\gamma$ line. The EWs can be measured only for the stars earlier than A0 spectral types. For the cooler stars it downs to 10 m\AA~which is below our measuring limit. The echelle spectra were extracted and wavelength calibrated by using a Fe-Ar lamp source with help of the IRAF {\sc echelle} package. 

  The selected classical Algol-type systems  are presented in Table 1. The apparent visual magnitudes and spectral types were collected from the SIMBAD data base and Ibano{\u g}lu et al. (2006).  Since the EWs of C species are depended on the effective temperature of the stars we observed some standard stars with the same instrumentation. The list of the standard stars and their properties are given in Table 2. Since the  C II $\lambda$ 4267 \AA~line is produced only for the effective temperatures higher than 10 000 K, the standard stars were selected from the main-sequence stars with the effective temperatures between 10 000 and 30 000 K.  
  
\begin{figure*}
\includegraphics[width=12cm]{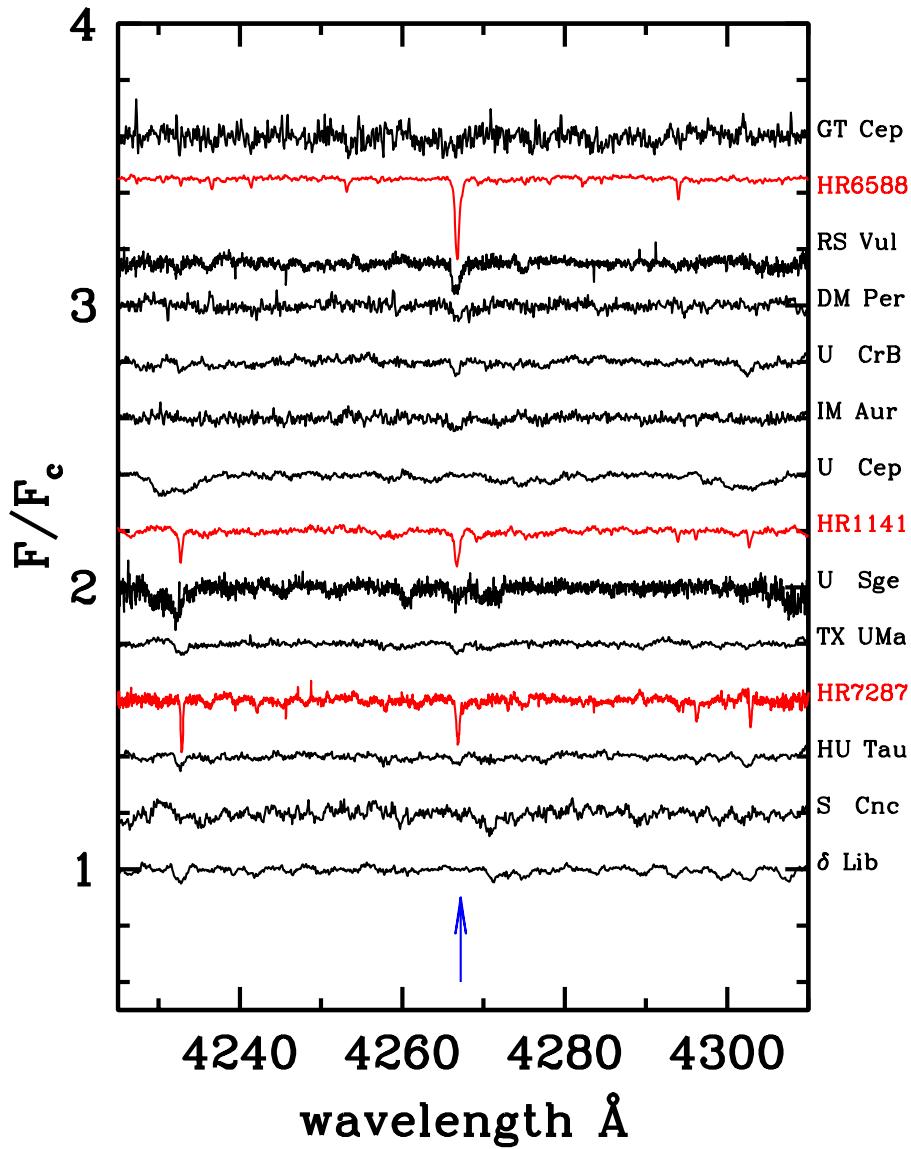}

\caption{Spectra of some Algols and standard stars covering the region of  C II $\lambda$ 4267 \AA, ordered with an increasing surface temperature from bottom to top. Wavelengths are corrected for orbital motion of the primary stars. The position of C II $\lambda$ 4267 \AA~line is also marked. }   
\label{Figure 1.}
\end{figure*}

\begin{table*}
\centering
%\begin{minipage}{125mm}
\caption{Observational details for the classical Algols. Visual magnitudes are taken from the SIMBAD database. The spectral types of the components are adopted from Ibano{\u g}lu et al. (2006). ASI and TUG refer to Asiago and Turkish National observatories, respectively. Signal-to-noise-ratios (S/N) were measured using the continuum at both sides of the C II $\lambda$ 4267 \AA~line. }
\label{Table 1.}
\begin{tabular}{lccccr}
\hline
Star	&V (mag)	&Spectral Type	&Observatory	&exposure time (s)	&S/N\\
\hline
IM Aur	&7.90	&B9 + K3 IV	&ASI	&2400	&80\\
S Cnc	&8.29	&B9.5V+G8-9 III-IV	&ASI	&2400	&80\\
U Cep	&6.75	&B7 V+G8 III-IV	&ASI	&1800	&95\\
GT Cep	&8.20	&B3 V+(A0)	&ASI	&2400	&50\\
U CrB	&7.66	&B6 V+F8 III-IV	&ASI	&2400	&120\\
V548 Cyg	&8.54	&A1 V+ F7	&ASI	&2400	&60\\
V1898 Cyg	&7.81	&B2IV + G2III	&TUG	&1200	&180\\
TW Dra	&8.00	&A8V+K0III	&ASI+TUG	&2400	&100 - 70\\
RY Gem	&8.69	&A2V+K0-3 IV-V	&ASI	&2400	&50\\
$\delta$ Lib	&4.91	&A0V+K0IV	&ASI	&800	&230\\
TU Mon	&9.00	&B5 V+F3	&ASI	&3600	&40\\
AU Mon	&8.11	&B5IV+F8-G0II-III	&ASI	&1800	&45\\
DM Per	&7.86	&B5 V+A5 III	&ASI	&2400	&60\\
U Sge	&6.45	&B7.5V+G4III-IV	&TUG	&2700	&80\\
HU Tau	&5.85	&B8 V+ F5 III-IV	&ASI	&800	&185\\
TX UMa	&7.06	&B8V+G0III-IV	&ASI	&1800	&170\\
Z Vul	&7.25	&B2V+B9V	&TUG	&2700	&75\\
RS Vul	&6.79	&B5 V+G0 III-IV	&TUG	&2700	&80\\

\hline \\
\end{tabular}
%\end{minipage}
\end{table*}

\begin{table*}
\centering
%\begin{minipage}{125mm}
\caption{Observational details of the standard stars. Visual magnitudes and spectral types are taken from the  SIMBAD database.}
\label{Table 2.}
\begin{tabular}{lccccr}
\hline
 Star	&V (mag)	&Spectral Type	&Observatory	&exposure time (s)	&S/N\\
\hline
HR1141	&5.65	&B6V	&ASI	&200	&150\\
HR1320	&4.28	&B3IV	&ASI	&300	&200\\
HR6588	&3.80	&B3IV	&ASI+TUG	&300	&250\\
HR7287	&5.14	&B8II	&TUG	&1200	&120\\
HR7426	&4.72	&B3IV	&TUG	&1200	&100\\

\hline \\
\end{tabular}
%\end{minipage}
\end{table*}

\section{Analysis}

\subsection{Variation of the C II $\lambda$ 4267 \AA~line equivalent widths }

The stellar continua of the stars having effective temperatures higher than 10 000 K are easily defined, in contrary to the cool stars, because of the existence of many spectral regions being relatively line-free at the neighbourhood  of the  C II $\lambda$ 4267 \AA~line. Therefore the equivalent width of this line can easily be measured in the spectra of the stars earlier than the spectral type of A0. As is presented in Tables 1 and 2 the S/N ratio varies with the apparent magnitudes of the stars. In Fig.1 we show a part of the spectra of some Algols and standard stars near the region of  C II $\lambda$ 4267 \AA~line. The spectra of the Algols are ordered according to their effective temperatures. As the effective temperatures lower the EWs of the C II $\lambda$ 426 7 line are also decreasing. The accuracy of an EW on a typical spectrum is estimated to be about 10 m\AA~which depends on the S/N ratio. 
Only one spectrum was taken for AI Dra, RW Gem and V356 Sgr.  The EWs of the  C II $\lambda$ 4267 \AA~line  could not be measured for these stars either insufficient exposure-times used or unsuitable orbital phases. In Table 3 we present EWs of the  C II $\lambda$ 4267 \AA~line for the standard stars measured by us and gathered from the previous studies. Since the meaning of any measured quantity depends on the associated uncertainty we calculated the errors of the measured EWs using the formula given by Cayrel (1988) and also by Stetson and Pancino (2008). They give an approximate formula for estimating the uncertainty of a measured EW as a function of $\Delta\lambda$ and signal-to-noise-ratio,

 \begin{equation}
\sigma_{ins} (EW) \simeq 1.6 \frac{\sqrt{\Delta\lambda \times EW}}{S/N}
\end{equation}

where $\Delta\lambda$ is the (constant) pixel size. This value has been taken as 0.086 and 0.036 for the spectra we obtained at Asiago and TUG observatories, respectively. Using the S/N ratios given in the last column of Table 1 the uncertainties are calculated and given in the fifth column of Table 4. Since the standard stars are very bright and, therefore, have higher S/N ratios their errors are about few milli-Angstroms (m\AA). Moreover we computed rms in m\AA~for stars with more than four spectra. The rms dispersion was listed in the last column of Table 4.

The effective temperatures for the primary stars are estimated, in general, from the wide-band B and V measurements of a system. Then, the effective temperatures for the secondary stars are obtained from the light curve analysis. In the literature different  $T_{eff}$ values are given for the same star which are not in agreement with its mass. In order to estimate the $T_{eff}$ values for the primary stars of the semi-detached Algol systems and the standard stars we decided to use a common temperature indicator as did by  Tomkin, Lambert and Lemke (1993). Since the strength or EW of the  C II $\lambda$ 4267 \AA~line is tightly depended on the effective temperature of the star we calculated the $T_{eff}$  values for the standard stars as well as the Algols using the intermediate-band photometric measurements. The [u-b] color and $ [c_{1}] $ index are very sensitive to the $T_{eff}$ of the stars and, therefore, they are effective temperature indicators for the stars earlier than A0. The calibration between $T_{eff}$ and [u-b] for a wide temperature interval, from 9500 to 30 000K, is adopted from Napiwotzki et al. (1993):

\begin{equation}
\Theta =  \frac{5040}{T} = -0.0195[u-b]^2 +0.2828[u-b]+ 0.01692
\end{equation}

On the other hand, we derived the following relationship between the $T_{eff}$ and $ [c_{1}] $ parameter using the $T_{eff}$ and $ [c_{1}] $-values given by Nissen (1974):
\begin{equation}
 \Theta = 0.3495(3)[c_1]+0.1696(2)
 \end{equation}
with a regression coefficient of 0.9991. This relationship is valid for the effective temperatures from 13 000 to 30 000 K.
The $T_{eff}$-values for the standard stars were computed using the the [u-b] and $ [c_{1}] $ parameters taken from Hauck and Mermilliod (1998, hereafter HM98)and are presented in Table 3. The EWs of C II $\lambda$ 4267 \AA~line for five standard stars are measured by us which are given in the first five lines of Table 3, the EWs for the others (including these stars) are taken from Hardorp and Scholz (1970), Kane et al. (1980), Kilian et al. (1989) and Tomkin, Lambert and Lemke (1993) . In Fig.2 we show the measured EWs/$\lambda$ of the standard stars as a function of the mean  effective temperatures of the stars computed by the Eqs. (2) and (3).  The variation of the EWs as a function of the $T_{eff}$ is represented by a third-order polynomial and shown by a continuous line. The EWs of $\lambda$ 4267 \AA~C II lines are increased up to 20 000 K and, then, decreased gradually.

\begin{table*}
\centering
%\begin{minipage}{85mm}
\caption{The [u-b] colours,$ [c_{1}] $ index, computed effective temperatures and the measured equivalent widths for the standard stars. }
\label{Table 3.}
\begin{tabular}{lccccr}
\hline
Star	&[u-b]	&$ [c_{1}] $	&$T_{eff}$ (K)	&EW (m\AA)	&Refs\\
\hline
HR1141	&0.683	&0.504	&14400	&103$\pm$1	&This Paper\\
HR6588	&0.424	&0.307	&17900	&205$\pm$5	&This Paper\\
HR1320	&0.558	&0.396	&16000	&152$\pm$6	&This Paper\\
HR7287	&0.793	&0.631	&13100	&84$\pm$14	&This Paper\\
HR7426	&0.543	&0.387	&16200	&157$\pm$7	&This Paper\\
HR39	&0.259	&0.137	&22000	&210	&1\\
HR153	&0.272	&0.152	&21600	&218	&3\\
HR811	&0.791	&0.614	&13200	&77	&3\\
HR922	&1.177	&0.951	&10300	&31	&3\\
HR1141	&0.683	&0.504	&14400	&106	&3\\
HR1320	&0.558	&0.396	&16000	&186	&3\\
HR1756	&0.041	&-0.039	&30100	&107	&4\\
HR1790	&0.227	&0.129	&22600	&239	&1\\
HR1855	&-0.018	&-0.073	&32800	&102	&2\\
HR1887	&0.046	&-0.035	&29800	&122	&2\\
HR2387	&0.069	&-0.003	&28300	&138	&1\\
HR2571	&0.093	&0.003	&27700	&186	&1\\
HR2806	&-0.050	&-0.101	&35000	&45	&2\\
HR2928	&0.177	&0.064	&24700	&146	&2\\
HR3023	&0.234	&0.099	&23100	&225	&2\\
HR3055	&-0.012	&-0.068	&32500	&118	&2\\
HR3468	&0.157	&0.063	&25000	&229	&2\\
HR5320	&0.085	&-0.022	&28600	&104	&2\\
HR6165	&-0.049	&-0.071	&33700	&82	&4\\
HR6588	&0.424	&0.307	&17900	&201	&3,1\\
HR7287	&0.793	&0.631	&13100	&74	&3\\
HR7426	&0.543	&0.387	&16200	&169	&3\\

\hline \\
\end{tabular}
%\end{minipage}
\\
(1.Kane et al. (1980), 2.Kilian et al. (1989), 3.Tomkin, Lambert and Lemke (1993), 4.Hardorp et al. (1970))

\end{table*}

The intermediate-band photometry of the Algol-type binaries were made by Hilditch and Hill (1975, hereafter HH75), Lacy (2002, hereafter CL02). The (\textit{b-y}) colors, $m_{1}  $,  $ c_{1} $ indices and $ H_{\beta} $ values were also given in the catalog of HM98. Since the orbital phases of the observations are known for the values obtained by HH75 and CL02 the parameters given by them are preferred to calculate the effective temperatures of the mass-gaining primary stars of the Algols. The observations obtained during the primary eclipse are excluded. However, it should be noted that the (\textit{b-y}) colors obtained by CL02 are systematically bluer than those obtained by HH75. If the stars were not observed by HH75 and CL02 we used the parameters given by HM98. The effective temperature computed for the primary of GT Cep is too high with respect to its mass, as given in Table 4. For this reason a $T_{eff}$ value of 19 000 K is adopted from Hohle et al. (2010).

The measured EWs are corrected using the formula given below:

\begin{equation}
\ EW_{0}= EW_{m} \frac{1+L}{L}
\end{equation}
where $ EW_{m} $ is the measured EW of the system and $ L $ is the mean continuum light ratio at the both side of the C II $\lambda$ 4267 \AA~line. The corrected $ EW_{0}/\lambda $ values for the primary stars of the Algols are also plotted in Fig.2 versus the effective temperatures, in logarithmic scale. Neither intermediate-  nor narrow-band photometric observations exist for the system V1898 Cyg, therefore, we adopted the effective temperature for the primary star from  Dervi{\c s}o{\u g}lu et al. (2011).

\begin{table*}
\centering
%\begin{minipage}{85mm}
\caption{The [u-b] colours, $ [c_{1}] $ index, computed effective temperatures, the number of measured EWs, the measured EWs, corrected EWs of the primary star, the uncertainty computed with Eq.(1) and standard deviations resulting from the measurements of the EWs.}
\label{Table 4.}
\begin{tabular}{lcccccccr}
\hline
Star	&[u-b]	&$ [c_{1}] $	&$T_{eff}$ (K)&n	&EW	(m\AA)&EW$_0$ (m\AA)	&$\sigma_{ins}$	&$\sigma_{SD}$\\
\hline
IM Aur		&0.626	&0.445	& 15200& 4	&74	&81	&2	&7\\
S Cnc		&1.140	&0.848	& 10800& 4	&14	&14	&1	&2\\
U Cep		&0.651	&0.469	& 14900& 3	&35	&40	&1	&3\\
GT Cep		&0.173	&0.081	& 24300$^*$& 5	&36	&45	&2	&4\\
U CrB		&0.612	&0.450	& 15200& 7	&72	&74	&1	&7\\
V548 Cyg	&1.182	&0.929	& 10300& 2	&14	&15	&1	&3\\
V1898 Cyg	&...	&...	& 18000& 28	&84	&86	&3	&17\\
TW Dra		&1.257	&0.788	& 10800& 6	&18	&19	&1	&5\\
RY Gem		&1.465	&1.132	& 9100 & 3	&8	&8	&1	&2\\
$\delta$ Lib		&1.208	&0.969	& 10200& 3	&40	&42	&0	&0\\
TU Mon		&0.416	&0.282	& 18300& 2	&72	&78	&3	&15\\
AU Mon		&0.457	&0.319	& 17500& 2	&34	&35	&2	&8\\
DM Per		&0.568	&0.420	& 15700& 4	&82	&87	&2	&15\\
U Sge		&0.734	&0.548	& 13900& 2	&97	&104	&	&30\\
HU Tau		&0.822	&0.598	& 13100& 3	&21	&22	&1	&10\\
TX UMa		&0.774	&0.563  & 13600& 8	&62	&64	&1	&12\\
Z Vul		&0.527	&0.388	& 16300& 2	&99	&111	&1	&20\\
RS Vul		&0.484	&0.370	& 16800& 4	&88	&91	&1	&9\\
\hline \\
\end{tabular}
%\end{minipage}
\end{table*}

\begin{figure*}
\includegraphics[width=12cm]{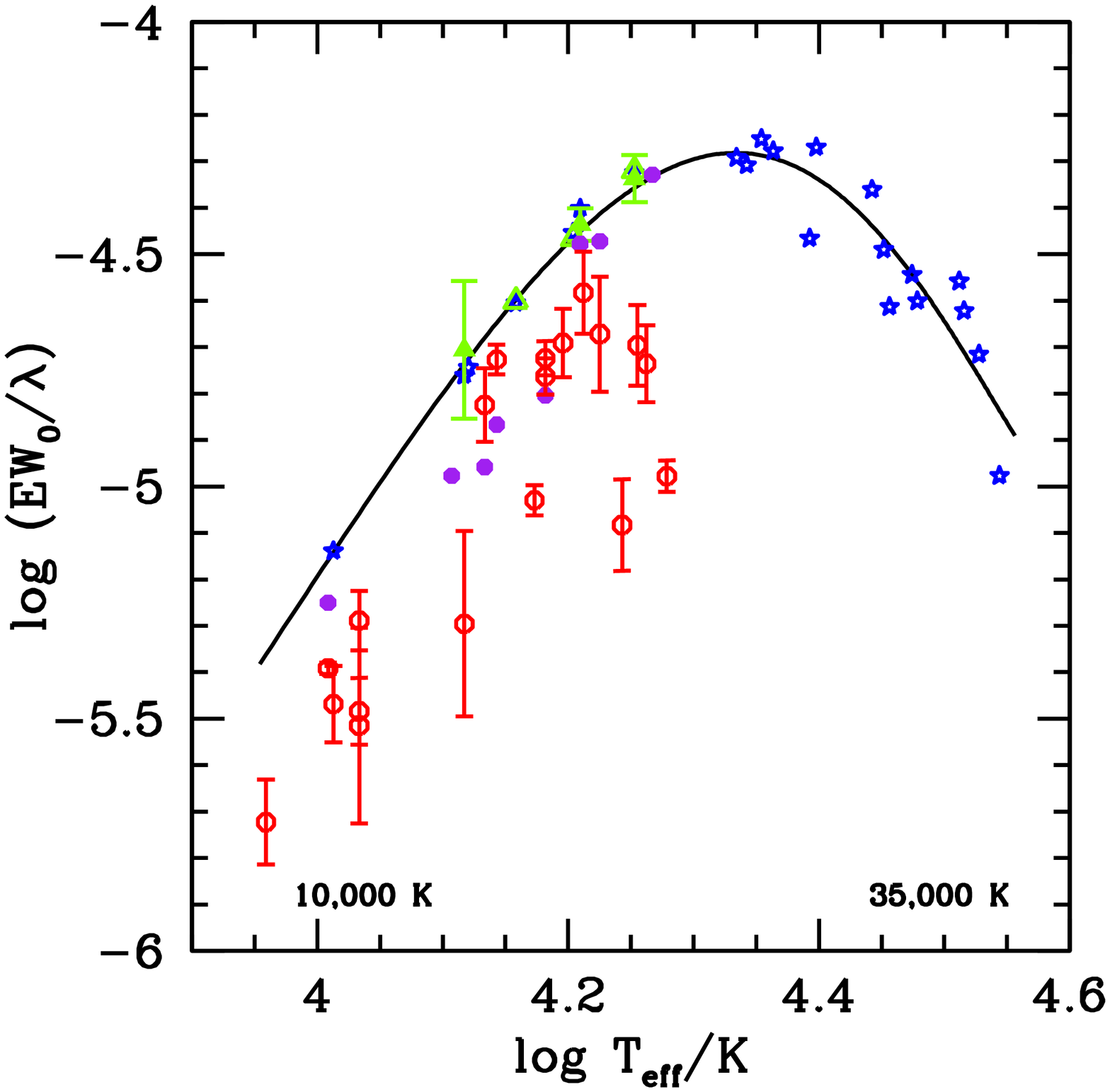}
\caption{The corrected EWs of the primary stars of the mass-exchanging Algols are plotted versus the effective temperatures. The symbols refer to: triangles and stars- standard stars observed by us and taken from literature; open circles- and dots- targets observed by us and Tomkin, Lambert and Lemke (1993), respectively. The error bars refer to the uncertainty resulting from the measurements of the EWs.}
\label{Figure 2.}
\end{figure*}

The EWs of $\lambda$4267 \AA~C II line in the gainers in the classical Algols are systematically smaller than that of the standard stars in the main-sequence band having similar effective temperatures. The weakest line is observed in the primary of AU Mon in which the EW is only 20 per cent that of the standard star with similar effective temperature. The weak line stars following the system AU Mon are GT Cep, HU Tau, U Cep and TU Mon in which the EWs are 22, 28, 34 and 40 per cent, respectively. 
We computed new-ODF ATLAS9 (Kurucz 1993) model atmospheres assuming a solar chemical composition ([M/H]=0.0), a microturbulent velocity $\xi$= 2 km/s, a surface gravity  $\log$\,g=4 and the effective temperatures given in Tables 3 and 4 for each star. 
Using a Linux version (Castelli, 2005) of the WIDTH-code(Kurucz 1993) and the corrected EWs we determined the carbon abundances for the primaries of classical Algols in our list. 
We used the line data from the NIST\footnote{\url{http://physics.nist.gov/cgi-bin/AtData/lines_form}}  database with the version 4, and adopted the log gf-values of +0.562 and +0.717 for this doublet. We added Stark damping constant $\log(\gamma_S/N_e)$ = -4.76 (Griem 1974) for all the stars. We determined the average C abundance for the five standard stars as $\log\,\varepsilon(C)=8.28 \pm0.10$ which is in a good agreement  with that given by Tomkin, Lambert and Lemke (1993) as 8.31 for the same stars.
  We take $\log\,\varepsilon(C)=8.52 $ for the solar abundance adopted from Grevesse and Sauval  (1998). When we compare with the solar value of 8.52 the standard stars have slightly smaller C abundance with an amount of 0.24 dex. The primary stars of Algols appear to have lower C abundances except $\delta$ Lib and RY Gem. The average C abundance of $\log\,\varepsilon(C)=7.75 \pm0.19$ is obtained for the 18 Algol primaries. On the other hand, the average $\log(N_{C} /N_{tot}) =-4.29 \pm0.22$ is obtained.   We estimate $ [N_{C} /N_{tot}] $ abundances of -1.91 for GT Cep, -1.88 for AU Mon and -1.41 for TU Mon (the poorer primaries), while +0.54 for $\delta$ Lib and +0.13 for RY Gem (richer primaries). The average  $ [N_{C} /N_{tot}] $ relative to the Sun is -0.82. The carbon abundances for the Algol primaries and standard stars are given in Table 5. In the last column of Table 5 we also present C abundances with respect to the Sun.  This result is clearly indicates that the primary stars contain much less carbon species compared to the main-sequence counterparts and also to the Sun.

\begin{table}
\centering
%\begin{minipage}{85mm}
\caption{The corrected EWs, the average abundance $\log\,\varepsilon(C)$, $\log\,(N_{C} /N_{tot})$ and the $ [N_{C} /N_{tot}] $ values with respect to average abundance, $\log\,\varepsilon(C)=8.52$, of the Sun.}
\label{Table 5.}
\begin{tabular}{lcccccccr}
\hline
Star	&EW$_0$ (m\AA)	&$\log\,\varepsilon(C)$ &$\log\,(N_{C} /N_{tot})$ &$ [N_{C} /N_{tot}] $	\\
\hline
IM Aur		&81	&7.80	& -4.24 	&-0.72	\\
S Cnc		&14	&7.99	& -4.05     & -0.53	\\
U Cep		&40	&7.28	& -4.76     & -1.24 \\
GT Cep		&45	&6.61	& -5.43     & -1.91	\\
U CrB		&74	&7.72	& -4.32     & -0.80	\\
V548 Cyg	&15	&8.27	& -3.77     & -0.25	\\
V1898 Cyg	&86	&7.26	& -4.78     & -1.26	\\
TW Dra		&19	&8.18	& -3.86     & -0.34	\\
RY Gem		&8	&8.65	& -3.39     & 0.13	\\
$\delta$ Lib&42	&9.06	& -2.98     & 0.54	\\
TU Mon		&78	&7.11	& -4.93     & -1.41	\\
AU Mon		&35	&6.64	& -5.40     & -1.88	\\
DM Per		&87	&7.74	& -4.30     & -0.78	\\
U Sge		&104&8.42   & -3.62     & -0.10	\\
HU Tau		&22	&7.40	& -4.64     & -1.12	\\
TX UMa		&64	&8.06   & -3.98     & -0.46	\\
Z Vul		&111&7.83	& -4.21     & -0.69	\\
RS Vul		&91	&7.53	& -4.51     & -0.99	\\

HR1141		&103&8.26	& -3.78     & -0.26	\\
HR1320		&152&8.25	& -3.79     & -0.27	\\
HR6588		&205&8.18	& -3.86     & -0.34	\\
HR7287		&84	&8.48	& -3.56     & -0.04	\\
HR7426		&157&8.22	& -3.82     & -0.30	\\

\hline \\
\end{tabular}
%\end{minipage}
\end{table}

\begin{figure*}
\includegraphics[width=8cm]{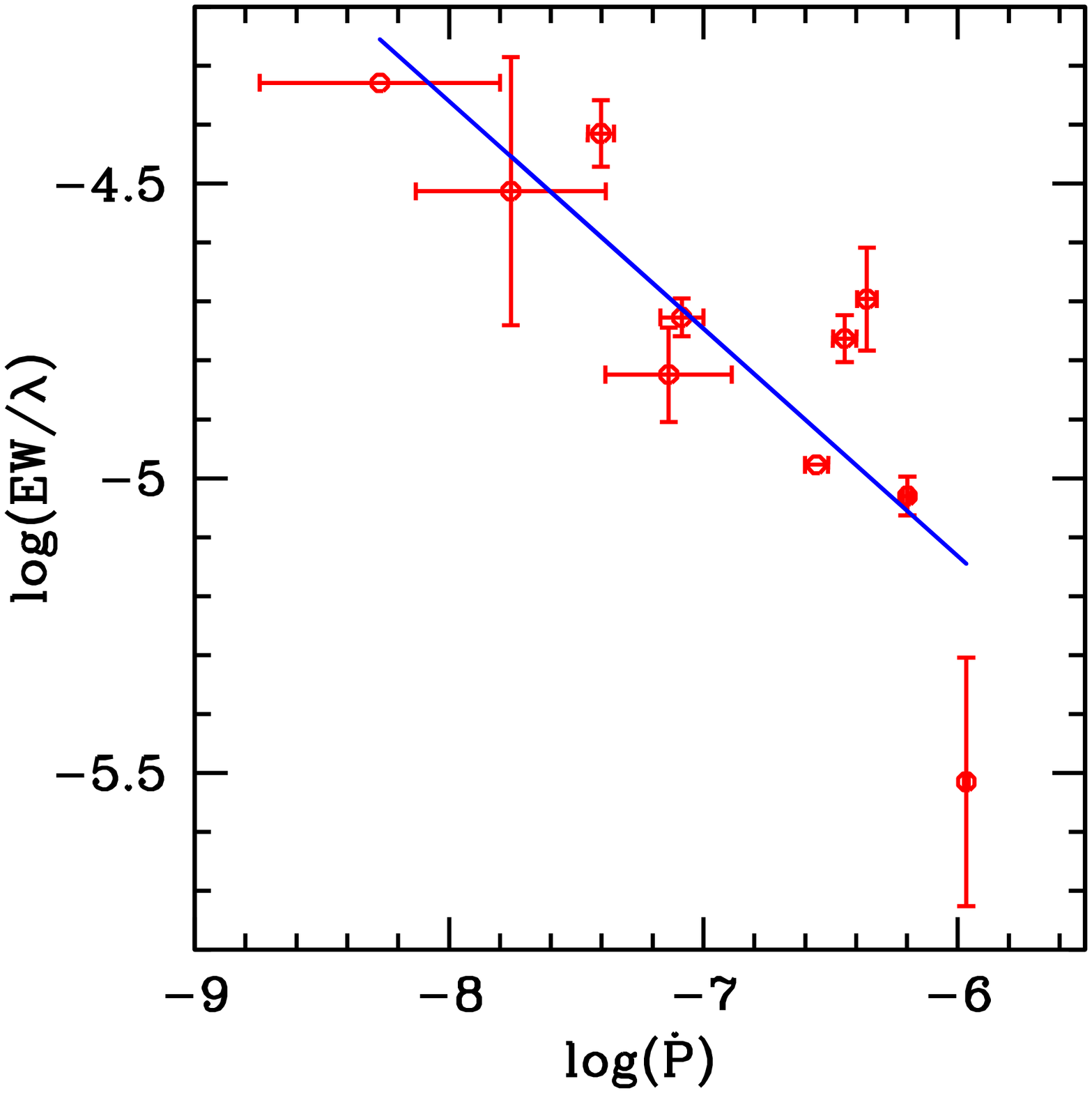}
\includegraphics[width=8cm]{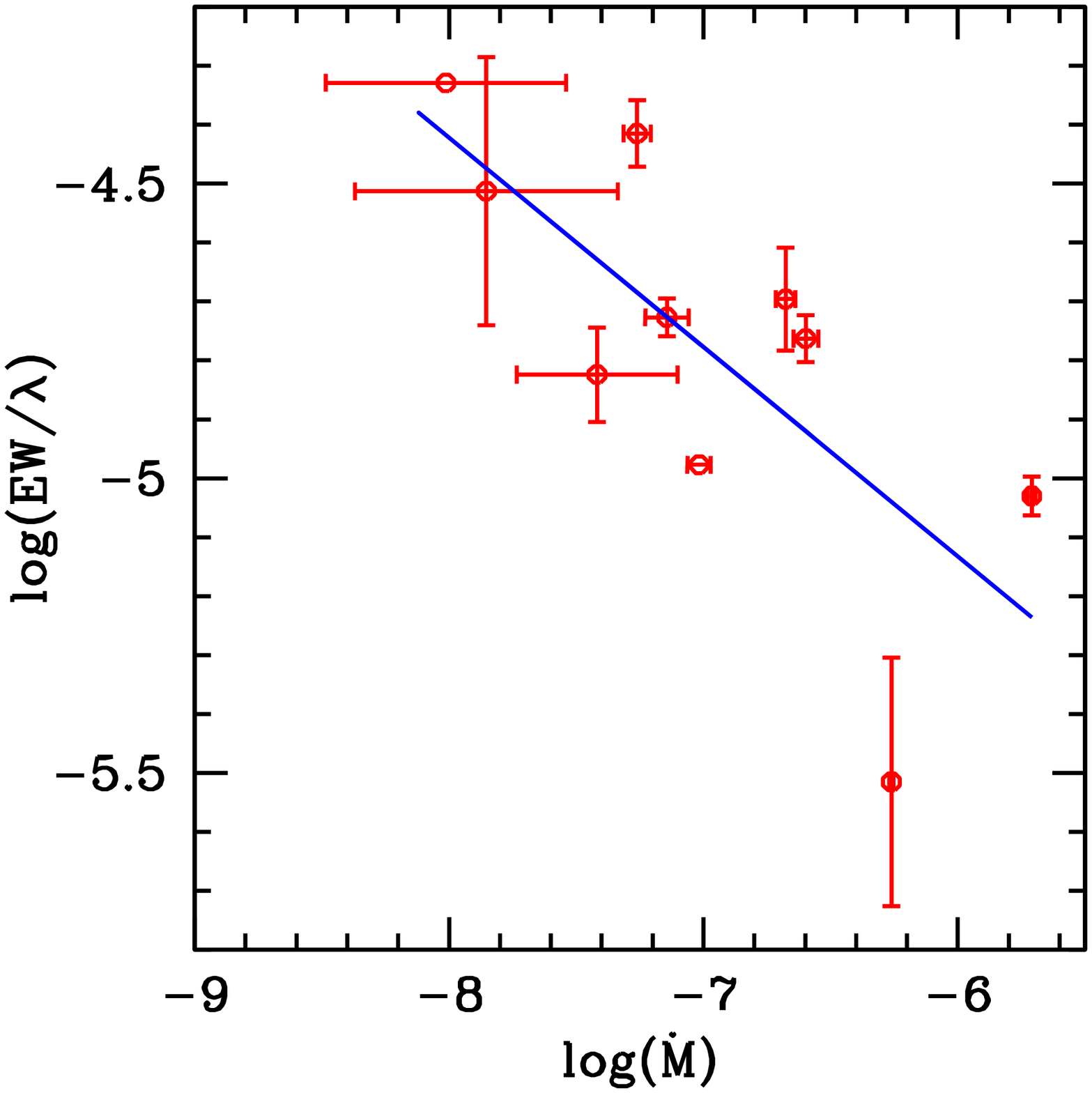}

\caption{The EWs are plotted as a function of the orbital period change rate in left. In the right panel EWs are plotted versus the mass transfer rate in the systems. A linear fit to the data is shown  
by the solid lines (see text).}
\label{Figure 3.}
\end{figure*}

\subsection{Period changes and mass transfer rates}

In most of the Algol-type binaries the orbital periods vary with time. The orbital period variations can easily be revealed  by the observations of mid-eclipses covering a sufficient time-interval. The behaviour of difference between the observed and computed times (O-C) as a function of epoch number (E) presents some hints why the orbital period of the system taken into account does change with time. As mentioned in the introduction the Algol-type variables are semi-detached interacting eclipsing binaries  in which less massive evolved components fill their critical Roche lobes and are transferring their material onto the more massive primary stars. The observed long-term  continuous increase of the orbital period may be attributed to the mass transfer from less massive evolved secondaries to the unevolved primaries, under the conditions of total mass and angular momentum conservation. However, most of the Algols show much more complicated orbital period changes than that expected from  the mass transfer. Some binaries display a periodic O-C variation superimposed on the period increase. The periodic part of the O-C variation caused by a third star. In some systems the orbital period tends to decrease with time which is indicative of the 
mass-loss from the system. Moreover, the losers are late type stars and, therefore, have magnetic activity which causes cyclic O-C variations. 

The times for mid-eclipses of eclipsing binary systems were listed in the two well-known extensive databases, i.e.,  Kreiner (2004) and Paschke and Brat (2006) O-C Gateway. All available photoelectric and CCD minima for the binaries, taken into account in this study, were collected from these databases and checked against the original literature. These databases include times for mid-eclipse obtained up to 2004 and 2006, therefore, we added some times of minima from the literature, published later. We used only photoelectric or CCD timings to obtain the rates of orbital period increase and mass transfer. In the case of TW Dra we used all the observed timings because it shows a very complicated O-C variation. In the ten systems the O-C variations can be represented by an upward parabola indicating orbital period increase, which is originated from mass transfer from less massive secondary to the more massive primary star.  The stars showing orbital period increase are listed in Table 6. The results of the linear least-squares solutions are listed in this table. The O-C analyses for the systems IM Aur, S Cnc, V548 Cyg, RY Gem, $\delta$ Lib and HU Tau indicate that the orbital periods are slightly decreasing, may be caused by (1) mass loss from the systems which overwhelms the effect of mass transfer, (2) third-body orbit, or (3) magnetic activity of the cooler star. In the cases of (2) and (3) a part of the sinusoidal change or cyclic variation may have been observed. In the three systems the orbital periods appear to almost constant.

\begin{table*}
\centering
%\begin{minipage}{125mm}
\caption{Results of the O-C analysis. The new ephemeris T$_0$, P, coefficient of the second order term (Q), number of observations, rates of the orbital period increase and mass transfer are listed, respectively, with standard deviations of each entry.   }

\label{Table 6.}
\begin{tabular}{lrrrrrr}
\hline
System 	&T$_0$ (JD 2 400 000+)	&P (days)&Q	&n	&$\dot{\rm{P}}$\, (yr$^{-1}$)	&$\dot{\rm{M}}$\, ($\rm{M_\odot\, yr^{-1}} $)\\
\hline
U Cep	&43753.7949(10)	&2.49306279(63)	&5.39(20)$\times 10^{-9}$	&193	&6.34(24)$ \times 10^{-7}$	&1.95(7)$ \times 10^{-6}$\\
U CrB	&44382.8522(26)	&3.4522186(14)	&5.86(62)$ \times 10^{-9}$	&40	&3.59(38)$ \times 10^{-8}$	&2.53(28)$ \times 10^{-7}$\\
V1898 Cyg	&50690.6938(8)	&1 .5131260(2)	&1.38(13)$ \times 10^{-9}$ 	&20	&4.39(39)$ \times 10^{-7}$	&2.11(19)$ \times 10^{-7}$\\
TW Dra	&34994.4350(49)	&2.80678014(84)	&1.165(24)$ \times 10^{-8}$	&531	&1.080(2)$ \times 10^{-6}$	&5.48(11)$ \times 10^{-7}$\\
U Her	&47611.5007(15)	&2.05102685(68)	&3.09(6.07)$ \times 10^{-11}$	&37	&5.35(5.82)$ \times 10^{-9}$	&0.97(1.10)$ \times 10^{-8}$\\
$\beta$ Per	&48275.1527(17)	&2.86732735(53)	&3.14(35)$ \times 10^{-9}$	&46	&2.78(29)$ \times 10^{-7}$	&9.62(1.02)$ \times 10^{-8}$\\
U Sge	&47390.3252(9)	&3.38061668(36)	&1.29(28)$ \times 10^{-9}$	&49	&8.22(1.61)$ \times 10^{-8}$	&7.19(1.41)$ \times 10^{-8}$\\
TX UMa	&46128.5347(33)	&3.0632855(18)	&9.38(4.85)$ \times 10^{-10}$	&72	&7.30(4.18)$ \times 10^{-8}$	&3.82(2.78)$ \times 10^{-8}$\\
Z Vul	&44852.5045(04)	&2.454931334(88)	&3.26(38)$ \times 10^{-10}$	&41	&3.95(46)$ \times 10^{-8}$	&5.49(68)$ \times 10^{-8}$\\
RS Vul	&45229.2983(12)	&4.47766471(49)	&4.80(3.34)$ \times 10^{-10}$	&16	&1.75(1.50)$ \times 10^{-8}$	&1.40(1.67)$ \times 10^{-8}$\\

\hline \\
\end{tabular}
%\end{minipage}
\end{table*}

In the last two columns of Table 6 the rates of the orbital change and mass transfer are given. While the highest mass transfer rate is found for U Cep, the smallest mass transfer rate is obtained for the system U Her. The EWs of $\lambda$4267 \AA~C II line in the 10 stars are plotted versus the orbital period changes and mass transfer rates in Fig.3, in logarithmic scale. The EWs for the systems U Her and $\beta$ Per are adopted from Tomkin, Lambert and Lemke (1993). It seems that there is a linear relationship between these two quantities. As the rate of mass transfer increases the  EW of $\lambda$4267 \AA~C II line decreases. De Greve (1993) suggests that only a fraction of the mass lost by the less massive companion is captured by the more massive primary star. As the orbital periods longer the matter captured by the primary decreases. The accreted material will be low during the fast phase of mass transfer, therefore the abundances of the gainer will not be influenced. The Algols studied here are in the phase of slow mass transfer. In this phase most of the material is captured by the primary star, therefore slow mass transfer can modify the atmosphere of the gainer. Since the transferring  material contains lower C and mixing within the original material the abundance of C appears to decrease (Sarna and de Greve, 1997). Fig.3 shows a correlation between the orbital period change and EWs of CII  $\lambda$4267 \AA~line of ten Algol-type interacting binary systems. Despite limited number of samples, there seems to be a correlation between the rates of orbital period increase and mass transfer and carbon abundance in the surface layers of the primary stars. As the rate of mass transfer increases the EW of C II line decreases.  

  \begin{figure*}
\includegraphics[width=10cm]{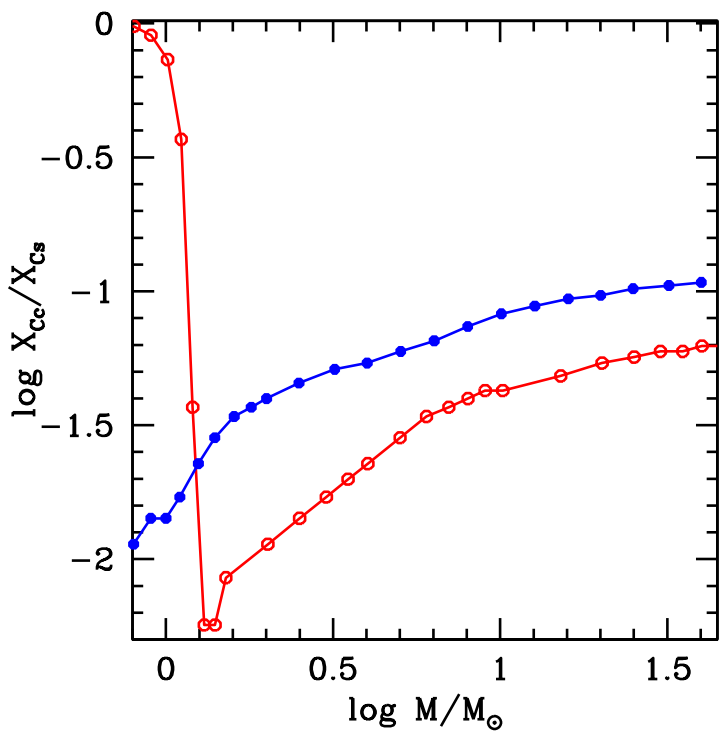}

\caption{Calculated ratio of center carbon to the surface value from ZAMS model (open circles) and from TAMS model (filled circles) according to Eggleton (1971).}
\label{Figure .}
\end{figure*}

\subsection{Evolutionary considerations}
As a point of single star evolution, stars that are massive than 1.4 \Msun \, fuse their hydrogen to helium by CNO-bicycle mechanism (Sarna and de Greve, 1997). As a net result of this process, while the the amount of carbon is reduced the nitrogen abundance is increased. Owing to nature of CNO cycle that occurs in the high-intermediate stars, the size of the reduced carbon regions reach approximately half of the star's mass due to the core convection at the end of main sequence. In Figure 4, we calculated the ratio of mass fraction of the carbon at the centre to the surface for zero-age main-sequence (ZAMS, open circles) and for terminal-age main-sequence (TAMS, solid circles) by using {\sc TWIN} stellar evolution code (Eggleton 1971) for a wide range of stellar masses.
	It is also known that the components of the binary stars spend their life as a single star until the originally more massive stars fill their corresponding Roche lobes. Following this stage of evolution the primary stars  transfer their mass at least up to initial mass ratio is reversed. The amount of material lost from losers slightly depends on whether the total AM is conserved ( Dervi{\c s}o{\u g}lu, 2010). As a result, the CNO processed material unavoidably is accreted onto the surface of gainers, originally less massive secondary stars. The amount of carbon reduced material on the surface of the today's primaries are expected as shown in Fig.4, where the ratios of C abundance as a mass of per gram matter in the core to the surface are plotted for the ZAMS and TAMS models. As it is clearly seen in Figure 4, the ratio of carbon at the core to the surface increases with  increasing mass. The ratio of C/N in the surface of primaries may give us some hints about the initial mass of the donor stars when fusing its hydrogen fuel. Therefore one may guess about the mass loss and accreted matter during the evolution of a close binary system.  

 \section{Conclusions}

 We have determined C abundance in the atmospheres of the primary stars of the 18 Algol-type systems. For determination of the C abundance we used the EWs of C II $\lambda$ 4267 \AA~doublet. A comparison of the C II $\lambda$ 4267 \AA~line with those of the standard stars, having the same effective temperature and luminosity class, clearly indicates that the gainers have significantly smaller EWs. The observed C abundances imply that the surface layers of the gainers have been altered by the accreted material which contains lower abundance of carbon than that in the atmosphere of this star. A plot of the EWs of C II $\lambda$ 4267 \AA~line versus the mass transfer rates for ten stars points out an existence of a  relationship between these quantities. The higher mass transfer rate corresponds to the smaller EWs. This may be taken as mixing on the surface layers of the accreted star which leads to lower C and higher N abundance (de Greve and Cugier, 1989). Logarithmic abundances relative to the standard stars and the Sun clearly confirm this expectation. Average $ [N_{C} /N_{tot}] $ are estimated to be -0.82 and -0.54 relative to the Sun and the standard stars, respectively. Specially, abundance ratios will give us important information about evolutionary histories and current status of the interacting Algols. However, it should be noted that measuring the isotopic ratios in the mass-losing secondaries is difficult due to their faintness, contributing a few per cent to the total light.   

\section{Acknowledgements}
The authors acknowledge generous allotments of observing time at Asiago Observatory (Italy) and TUBITAK National Observatory (TUG) of Turkey. We also wish to thank the Turkish Scientific and Technical Research Council for supporting this work through grant No. 109T708 and T\"UB\.ITAK National Observatory (TUG) for a partial support in using 
RTT150 with project number  10ARTT150-493-0.

\end{document}